\documentclass[12pt,draftclsnofoot,onecolumn]{IEEEtran} \usepackage{amssymb}
\usepackage{amsmath}
\usepackage{graphicx}
\usepackage{cite}
\usepackage{citesort}
\usepackage{color}	
\usepackage{subfigure}
\usepackage{graphicx,epstopdf}
\usepackage{epsfig}	
\usepackage{cite,graphicx,amsmath,amssymb}
\usepackage{comment}
\usepackage{bm}
\usepackage{lipsum}
\usepackage{stfloats}
\usepackage[ruled,linesnumbered]{algorithm2e}

\newtheorem{theorem}{\textbf{Theorem}}

\newtheorem{lemma}{\textbf{Lemma}}

\newtheorem{corollary}{\textbf{Corollary}}

\makeatletter
\def\ScaleIfNeeded{%
\ifdim\Gin@nat@width>\linewidth \linewidth \else \Gin@nat@width
\fi } \makeatother

\begin{document}
%

\title{\huge{Spatial-spectral Cell-free Networks: A Large-scale Case Study}}
\author{Zesheng Zhu, Lifeng Wang, Xin Wang,  Dongming Wang and Kai-Kit Wong
\thanks{Z. Zhu, L. Wang, X. Wang are with the School of Information Science and Engineering, Fudan University, Shanghai 200433, China (e-mail: $\rm\{lifengwang,xwang11\}@fudan.edu.cn$).}
\thanks{Dongming Wang is with the National Mobile Communications Research Laboratory,
Southeast University, Nanjing 210096, China (e-mail: wangdm@seu.edu.cn).}
\thanks{K.-K. Wong is with the Department of Electronic and Electrical Engineering, University College London, London WC1E 7JE, U.K.; K.-K. Wong is also affiliated with Yonsei Frontier Lab, Yonsei University, Korea  (e-mail:
kai-kit.wong@ucl.ac.uk).}
}

\maketitle

\begin{abstract}
This paper studies the large-scale cell-free networks where dense distributed  access points (APs) serve many users. As a promising next-generation network architecture, cell-free networks enable ultra-reliable connections and minimal fading/blockage, which are much favorable to the millimeter wave and Terahertz transmissions. However,  conventional beam management with large phased arrays in a cell is very time-consuming in the higher-frequencies, and could be worsened when deploying a large number of coordinated APs in the cell-free systems. To tackle this challenge, the spatial-spectral cell-free networks with the leaky-wave antennas are established by coupling the propagation angles with frequencies. The beam training overhead  in this direction can be significantly reduced through exploiting such spatial-spectral coupling effects. In the considered large-scale spatial-spectral cell-free networks, a novel subchannel allocation solution at sub-terahertz bands is proposed by leveraging the relationship between cross-entropy method and mixture model. Since initial access and AP clustering play a key role in achieving scalable large-scale cell-free networks, a hierarchical AP clustering solution is proposed to make the joint initial access and cluster formation, which is adaptive and  has no need to initialize the number of AP clusters. After AP clustering, a  subchannel allocation solution is devised to manage the interference between AP clusters. Numerical results are presented to confirm the efficiency of the proposed solutions and indicate that besides subchannel allocation, AP clustering can also have a big impact on the large-scale cell-free network performance at sub-terahertz bands.
\end{abstract}

\section{Introduction}
Compared with the 5G massive multiple-input multiple-output (MIMO), cell-free massive MIMO system can significantly improve the spectral efficiency and system scalability~\cite{Ngo2017,Giovanni2019,EmilBJ2020}. Recent measurements~\cite{Arnold2021} demonstrate that cell-free massive MIMO provides ultra-reliable connections and minimizes the fading/blockage, which are of the utmost importance at 6G millimeter wave (mmWave)/terahertz(THz) bands. Therefore, cell-free network is envisioned as the next-generation wireless network architecture~\cite{chengxiangWang}.

Existing research contributions have paid much attention to the cell-free massive MIMO systems in the lower-frequencies~\cite{Elina2015,trung2018,Bashar2019,Karlsson2020,Chongzheng2024,Jianxin2024}.
Early work~\cite{Elina2015} has made the spectral efficiency comparison between the cell-free massive MIMO systems and small cell networks and shown that
more than ten-fold improvement on each user's data rate can be achieved.
 A closed-form expression for the spectral efficiency with channel estimation errors and power control concerns is derived in~\cite{trung2018}, to theoretically affirm that compared to the colocated counterpart, cell-free massive MIMO systems enable uniformly good services and higher energy efficiency. Joint receiver filter and uplink power control design is studied in \cite{Bashar2019}, where the proposed scheme aims to maximize the minimum uplink user rate. To reduce the channel training cost,  local partial zero-forcing and local protective
partial zero-forcing precoding schemes for cell-free massive MIMO are developed in~\cite{Karlsson2020}. The work of \cite{Chongzheng2024} focuses on the joint user association and power control in a cell-free massive MIMO system, to maximize the sum spectral efficiency. Latest work \cite{Jianxin2024} introduces the reconfigurable intelligent surface aided cell-free massive MIMO systems, to further mitigate the fading effects and enhance the energy efficiency.

Since links in the higher-frequencies suffer high path loss and the blockage, the use of large mmWave and THz frequency bands in 5G/6G creates great potential for cell-free massive MIMO.  Therefore, the advantages of cell-free massive MIMO at mmWave and THz bands have been exploited in the literature~\cite{Alonzo2019,Seungnyun2021,WaiShuki2023,Qinyuan2024,junhui2024,Buzzi2021mobile,Zaher2024}.  Power control for cell-free mmWave downlink and uplink have been designed to maximize the energy efficiency in \cite{Alonzo2019}, where no channel estimation is needed at the mobile stations. To reduce the channel feedback overhead, \cite{Seungnyun2021} selects a few dominating paths for cell-free mmWave dowlink precoding due to the sparse scattering environment at mmWave. The beam search issue for cell-free mmWave links has to be addressed for low-latency transmissions, thus  \cite{WaiShuki2023} proposes a fast beam search design based on the prior beam combinations information. In \cite{Qinyuan2024},   the cell-free mmWave power model including transmit power, circuit power and load-dependent backhaul power is built, and joint optimization of user-access point (AP) association, hybrid beamformers and AP
selection is developed to enhance the energy efficiency. Meanwhile, \cite{junhui2024} provides a machine learning based scheduling scheme in mmWave cell-free vehicle-to-infrastructure  networks, to minimize the power consumption. The impacts of mobility and handover on the mmWave cell-free networks are investigated in \cite{Buzzi2021mobile,Zaher2024}, to enhance the scalability and robustness. However, beam management and link budgets at mmWave and THz bands pose great challenges to both cellular and cell-free networks with conventional phased arrays~\cite{Rikkinen_THz_2020}, and the initial access and beam scanning procedures are heavily time-consuming in realistic networks~\cite{WaiShuki2023,PoleseMag}.

In the cell-free mmWave and THz networks with conventional phased arrays, each AP has to select the optimal beam for one user equipment (UE) position, and the beam training complexity may scale linearly with the number of APs when using conventional all-AP exhaustive beam search~\cite{WaiShuki2023,PoleseMag}, which hinders the large-scale cell-free deployment and constrains the number of served UEs. Moreover, obtaining the channel state information of each transmit antenna element becomes unlikely in sub-THz/THz networks, and conventional antenna solutions based on such channel state information are not applicable~\cite{Lin_chen2015,Yasaman_2020}. Spatial-spectral mmWave/THz transmission is a  promising  way to substantially reduce the initial access and beam scanning time delay through coupling the propagation angles with frequencies~\cite{A_J_Seeds1995,D_headland2018,Linglong2022,Li_Ruifu2022,Ratnam2022,FDM_2015,LZ2022}. There are two important antenna solutions for making spatial-spectral mmWave/THz links, namely true-time delay (TTD) based phased array~\cite{A_J_Seeds1995,D_headland2018,Linglong2022,Li_Ruifu2022} and leaky-wave antenna~\cite{Yasaman_2020,D_headland2018,FDM_2015,LZ2022}. TTD  based phased array solution adopts the variable delay lines to enable that conventional phased array can also create frequency-dependent beams, and the delay range of the circuit blocks needs to be delicately chosen~\cite{D_headland2018,Li_Ruifu2022}, which may be more stringently required in cell-free massive MIMO systems. Leaky-wave antenna solution provides a low-cost alternative  since such an easy-to-manufacture traveling-wave antenna inherently enables spatial-spectral mmWave/THz waves. Therefore, leaky-wave antenna solution has been leveraged in frequency-division multiplexing THz communications~\cite{FDM_2015,Yasaman_2020}, sensing and tracking~\cite{Saeidi2021,LZ20222023}, and physical layer security~\cite{Chia-Yi_TFS}. The work of \cite{LZ_cl_2023}  presents a novel conditional generative adversarial networks (GAN) to estimate the channel of a THz link with leaky-wave antenna, which only requires single shot channel discovery. Therefore, by supporting the spatial-spectral transmissions, the hurdle of initial access and beam management in cell-free mmWave and THz networks may be overcome.

Motivated by the aforementioned works, this paper attempts to study the large-scale cell-free networks at sub-THz bands, where leaky-wave antennas are leveraged to make spatial-spectral links and dense APs simultaneously serve multiple UEs. The initial work~\cite{Zesheng2024} considers a single-user case and provides antenna selection schemes in a cell-free system with leaky-wave antennas, which shall be extended. Therefore, we focus on the generalized scenario and highlight the novel contributions as follows.
\begin{itemize}
  \item In the large-scale cell-free networks with leaky-wave antennas, the multiuser massive MIMO downlink transmission model is established. Differing from the conventional phased array case~\cite{Alonzo2019,Seungnyun2021,WaiShuki2023,Qinyuan2024,junhui2024,Buzzi2021mobile,Zaher2024}, the considered channel model is frequency-dependent and inherently reconfigurable thanks to the leaky-wave antenna's spatial-spectral property. As such, the precoding matrix is reliant on the frequency allocation.
  \item We first study the large-scale spatial-spectral cell-free system without AP clustering (namely one central processing unit case). In this case, frequency allocation problem is formulated to maximize the total downlink transmission rate. To deal with it, the formulated problem is translated into an estimation problem in a manner based on cross-entropy method. To avoid being stuck in the local minima, a Gaussian mixture model and its adaptive one are adopted to find better solution.
  \item Since AP clustering enhances the scalability  and it is more practical that UEs are served by a subset of APs in the large-scale cell-free networks~\cite{Ranjbar2022}, we further strive  to develop the scalable large-scale spatial-spectral cell-free system with AP clustering. Differing from the existing clustering schemes~\cite{EmilBJ2020,Yanlin2019,Michal2021,JTAO2023,Junyuan2023}, our AP clustering criterion is based on maximizing the spectral efficiency per AP in each cluster~\cite{Wagner2012} and a fast initial access approach with one-shot discovery is adopted. As such, a hierarchical AP clustering is proposed without the need to initialize the number of AP clusters, to address the impacts of both AP positions and access links on the AP clustering.  Furthermore, a  subchannel allocation solution is presented to avoid the co-channel interference among the AP clusters.
\end{itemize}
Numerical results are presented to confirm the efficiency of the proposed subchannel allocation and  hierarchical clustering algorithms.

The rest of the paper is organized as follows: Section II
outlines the system model. Section III proposes a subchannel allocation algorithm in the large-scale spatial-spectral cell-free system. Section IV designs the joint initial access and cluster formation. Section V presents numerical results and the concluding remarks are given in Section VI.

\text{\emph{Notations:}} In this paper, $\mathrm{sinc}\left(x\right) = \mathrm{sin} \left(x\right)/x$; $\left(\cdot\right)^T$ and  $\left(\cdot\right)^H$ are the transpose and conjugate transpose operators, respectively; $\odot$ is the Hadamard product operator;  $\left\|\cdot\right\|$ is the  $\rm L_2$ vector norm; $\left(\cdot\right)^{-1}$ is the inverse of a matrix; $\emptyset$ denotes the empty
set; ${\rm dim(\cdot)}$ denotes the number of entries for a vector; $\ln(\cdot)$ denotes the  natural logarithm. The cardinality of a set is denoted by ${\rm card}\left(\cdot\right)$.

\section{System Model}

\begin{figure}[t]
\centering
\includegraphics[width=3.2 in]{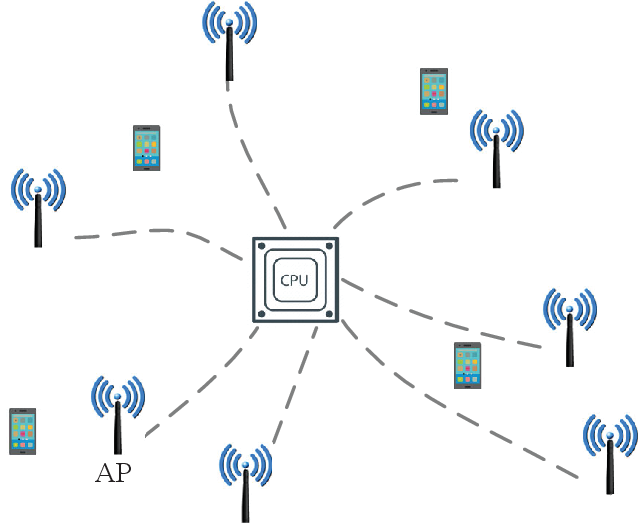}
\caption{An illustration of cell-free system.}
\label{cell_free}
\end{figure}

As illustrated in~Fig.~\ref{cell_free}, a cell-free system consists of many APs located in a distributed manner. The APs are connected to the central processing unit (CPU) via fronthaul links and CPU  executes the joint baseband processing. Each AP is equipped with one multi-face leaky-wave antenna for 360-degree coverage, and the effective leaky-wave antenna gain is~\cite{Sutinjo2008,Security2020}
\begin{align}\label{radiation}
 G\left(f,\theta\right) = \xi L{\rm{sinc}}\left[ {\left( { - \mathrm{j}\varrho  - {k_0}\left( f\right)\cos \theta  + \beta \left( f \right)} \right)\frac{L}{2}} \right]
\end{align}
where $\xi$ is the radiation efficiency factor, $L$ is the aperture length, $\mathrm{j}=\sqrt{-1}$, $\varrho$ is the attenuation coefficient, and $k_0\left(f\right)=2\pi f/c$, $\beta\left(f\right) = k_0\left(f\right) \sqrt{1-\left(\frac{f_{\rm co}}{f}\right)^2}$, $f$ denotes the subchannel center frequency, $f_{\rm co}$ is the cutoff frequency, and  $\theta$ is the  propagation angle.
Assuming that there are $M$ APs and $K$ active  single-antenna UEs in a spatial-spectral cell-free system, the downlink multiuser channel matrix  is given by
\begin{align}\label{channel_MIMO}
&{\bf{\tilde H}}\left(f\right) =\left[\widetilde{\mathbf{h}}_{1} \cdots \widetilde{\mathbf{h}}_{k} \cdots \widetilde{\mathbf{h}}_{K}\right]^T \nonumber\\
&= \left[ { \mathbf{H} \odot \left( {\begin{array}{*{20}c}
   {\sqrt {G\left( {f,\theta _{11} } \right)} } &  \cdots  & {\sqrt {G\left( {f,\theta _{1M} } \right)} }  \\
    \vdots  &  \ddots  &  \vdots   \\
   {\sqrt {G\left( {f,\theta _{K1} } \right)} } &  \cdots  & {\sqrt {G\left( {f,\theta _{KM} } \right)} }  \\
\end{array}} \right)} \right],
\end{align}
where $\mathbf{H}=\left[\mathbf{h}_{1} \cdots \mathbf{h}_{k} \cdots \mathbf{h}_{K}\right]^T \in \mathcal{C}^{K \times M} $ is the conventional channel matrix with omnidirectional antenna, $\mathbf{h}_{k}=[h_{1k},\cdots,h_{Mk}]^T \in \mathcal{C}^{M \times 1}$ is the downlink channel coefficient vector between the $k$-th UE and the $m$-th AP. It is indicated from \eqref{channel_MIMO} that the considered spatial-spectral cell-free transmissions make the channel reconfigurable, since the equivalent channel gains can be flexibly tuned by selecting the proper frequencies. Therefore,  spatial-spectral cell-free system has the potential to achieve more diversity gains than the existing cell-free solutions.

The received signal-to-interference-plus-noise ratio (SINR) at the $k$-th UE for the downlink transmission is calculated as
\begin{align}\label{SINR}
\gamma _{k}\left(f\right)= \frac{q_k \left|{\widetilde{\mathbf{h}}_k}^T \mathbf{w}_k \right|^2}{\sum\limits_{k^{'}  = 1,k^{'} \ne k}^K {q_{k^{'} } \left| {\widetilde{\mathbf{h}}_k}^T \mathbf{w}_{k^{'} }\right|^2 }  + \delta ^2 },
\end{align}
where $q_k$ is the transmit power spectral density (PSD) for the $k$-th UE's data signal, $\delta ^2$ is the noise's PSD, $\mathbf{w}_k=\frac{\mathbf{F}\left(:;k\right)}{\left\| \mathbf{F}\left(:;k\right)\right\|}$ is the normalized $k$-th column vector of the precoding matrix, when leveraging the maximum ratio transmission (MRT) and zero-forcing (ZF) precoding methods,  $\mathbf{F}$ is
\begin{align}\label{precoding_matrix}
\mathbf{F }= \left\{ \begin{array}{l}
{\bf{\tilde H}}^H\quad \quad \quad\quad\quad\quad~~ \;\,\mathrm{for\,MRT}, \\
{\bf{\tilde H}}^H \left( {\bf{\tilde H}} {\bf{\tilde H}}^H  \right)^{ - 1} \quad \quad \mathrm{for\,ZF}. \\
 \end{array} \right.
\end{align}
Thus, based on \eqref{SINR}, the total downlink transmission rate in a spatial-spectral cell-free system is given by
\begin{align}\label{DRT}
R_{\rm total}= \sum\limits_i {B_{i} } \sum\limits_{k = 1}^K \log_2\left(1+\gamma _{k}\left(f_i\right)\right),
\end{align}
where $B_{i}$ is the $i$-th subchannel bandwidth with the center frequency $f_i$.

\textbf{\emph{Remark 1}:} In the  mmWave or THz cell-free system with conventional antenna arrays, each distributed AP has to build the initial access through beam scanning~\cite{linchen2016,PoleseMag}, which is very time-consuming for the case of dense APs or UEs.  True time delay (TDD) based phased array is also able to create spatial-spectral links~\cite{Linglong2022,Li_Ruifu2022}, however, in such a design, variable delay lines shall be delicately employed to cover the angular space for all the UEs, in addition, high-resolution spatial direction  requires large numbers of subcarriers in the TDD based spatial-spectral system, which may bring in high PAPR~\cite{Seung2005}. As indicated from \eqref{channel_MIMO}, the cost-effective leaky-wave antenna inherently enables spatial-spectral coupling beams~\cite{D_headland2018}, which makes the cell-free system more scalable.

\section{Subchannel Allocation in Large-scale Cell-free System without AP Clustering}\label{SA-CF}
In this section, we focus on the subchannel allocation in the cell-free system where a large number of APs serve multiple UEs at the same time and frequency band. Differing from the conventional counterpart, spatial-spectral cell-free system has frequency-dependent beams. Therefore, the subchannel allocation needs to be re-designed.  To maximize the transmission rate, the subchannel allocation problem is formulated as
\begin{align}\label{Opt_problem}
&\mathop {\max }\limits_{{\bf{B,f}}} R_{\rm total} = \sum\limits_i {B_{i} } \sum\limits_{k = 1}^K \log_2\left(1+\gamma _{k}\left(f_i\right)\right)\\
&\mathrm{s.t.} ~\mathrm{C1:}~\sum\limits_i {B_i }  \le B_{\rm total}, \nonumber\\
&\mathrm{C2:}~\bigcap\limits_{i} {\left\{ {f\left| {f \in \left[ {f_{i}  - \frac{{B_{i} }}{2},f_{i}  + \frac{{B_{i} }}{2}} \right]} \right.} \right\}}  = \emptyset , \nonumber\\
&\mathrm{C3:}~ \gamma _{k}\left(f_i\right) \geq {\bar{\gamma}_{\rm th}}, \forall k, \forall i, \nonumber \\
&\mathrm{C4:}~\left\| {{\widetilde{\gamma}_k} \left( {f_{i}  - \frac{{B_{i} }}{2}} \right)\left| {_{\rm dB} } \right. - {\widetilde{\gamma}_k} \left( {f_{i}  + \frac{{B_{i} }}{2}} \right)}\left| {_{\rm dB} } \right. \right\| < \varepsilon , \forall k, \forall i, \nonumber \\
&\mathrm{C5:}~B_i \geq 0, \;~ f_{i} \geq 0, \;\,\;\forall i, \nonumber
\end{align}
where $\mathbf{B}=[B_i]$ and $\mathbf{f}=[f_i]$ are the subchannel bandwidth vector and center frequency vector, respectively; ${\widetilde{\gamma}_k}\left( f_i\right)$ denotes the $k$-th UE's received signal power for the $i$-th subchannel. Constraint $\mathrm{C1}$ is the frequency bandwidth limitation with the maximum value $B_{\rm total}$; $\mathrm{C2}$ ensures that there is no overlap between subchannels; $\mathrm{C3}$ is the minimum receive power constraint with the threshold ${\bar{\gamma}_{\rm th}}$; $\mathrm{C4}$ guarantees that  the received signal strength (RSS) difference in the frequencies of a subchannel is less than a small value $\varepsilon$, which is also referred to as the coherence bandwidth of a subchannel. Note that each AP has the same subchannels with the same center frequencies in the cell-free system.

The problem \eqref{Opt_problem} is non-differentiable with non-overlap constraint (combinatorial).  Although a closed-form subchannel allocation solution for the single-input single-output case is obtained in~\cite{LZ2022}, such a heuristic  method is infeasible in the considered multiuser cell-free system. In this work, we develop a cross-entropy based algorithm to solve the problem \eqref{Opt_problem}. Cross-entropy method is a powerful tool to adaptively solve the combinatorial and continuous problems~\cite{Kochenderfer2019}, particularly large non-convex problem~\cite{AmosBrandon_2020}. Its gist is that optimization problems can be translated into the estimation problem in a stochastic manner, which has been substantially  applied in many areas such as action components for end-to-end machine learning pipeline~\cite{AmosBrandon_2020}, device placement~\cite{Yuanxiang2018}, planning~\cite{ZhangZichen2022}, and safe reinforcement learning~\cite{Weimin2021}, etc.

Since the subchannel center frequency  and its corresponding subchannel bandwidth are coupled, we first suppose that the subchannels' center frequencies follow  a Gaussian mixture model (GMM), i.e., the $i$-th subchannel center frequency $f_i\sim p\left(f|{\bm \varpi},{\bm \mu},{\bm \vartheta}\right)$, where the GMM $p\left(f|{\bm \varpi}_i,{\bm \mu}_i,{\bm \vartheta}_i\right)$ is defined as
\begin{align}\label{GMM}
p\left(f|{\bm \varpi},{\bm \mu},{\bm \vartheta}\right)= \sum\limits_{\ell  = 1}^{\varsigma} {\varpi _{\ell } } \mathcal{N}\left(f| \mu_{\ell },\vartheta_{\ell }\right),
\end{align}
where ${\bm \varpi}=[\varpi _{\ell } ]$ is the mixture weight vector, ${\bm \mu}=[\mu_{\ell }]$ and ${\bm \vartheta}=[\vartheta_{\ell }]$ are the mean and variance vectors of the mixture components, respectively, and $\mathcal{N}\left(f| \mu_{\ell },\vartheta_{\ell }\right)$ denotes the $\ell$-th mixture component following a Gaussian distribution. At each iteration, there are $N_{\rm c}$ candidate samples $\{f_i^{1},\cdots,f_i^{N_{\rm c}}\}$ generated by the GMM $p\left(f|{\bm \varpi},{\bm \mu},{\bm \vartheta}\right)$ for the $i$-th subchannel center frequency. Given the candidates $\{f_i\}$, the corresponding subchannel bandwidth  $\{B_i\}$ can be obtained by solving the problem \eqref{Opt_problem} via one-dimensional search, and thus the reward (objective) $R_{\rm total}$ is calculated. One important hyperparameter of the cross-entropy method is that $N_{\rm elite}$ ($N_{\rm elite}<N_{\rm c}$) elite samples $\{f_i^{(1)},\cdots,f_i^{(N_{\rm c})}\}$ with larger reward values are chosen among the candidate samples, i.e., $f_i^{(n)}$ is the $i$-th subchannel center frequency sample with the $n$-th largest reward. Based on these elite samples, the expectation-maximization (EM) algorithm is adopted to determine the GMM. Specifically, by alternating E-step with M-step in the EM procedure, the GMM parameters are estimated. In light of Baye's rule, the E-step is given by
\begin{align}\label{GMM-E}
{\rm \textbf{E-step:}} \;\varphi _i^{\ell ,n}  = \frac{\varpi _{\ell }\mathcal{N}\left(f_i^{(n)}| \mu_{\ell },\vartheta_{\ell }\right)}{{\sum\limits_{\ell^{'} = 1}^\varsigma \varpi _{\ell^{'} } {\mathcal{N}\left(f_i^{(n)}| \mu_{\ell^{'} },\vartheta_{\ell^{'}  }\right)} }},
\end{align}
where $\varphi _i^{\ell ,n}$ is the posterior probability of elite sample $f_i^{(n)}$ belonging to the $\ell$-th mixture component. Then, GMM parameters are estimated at M-step, i.e.,
\begin{align}
&{\rm \textbf{M-step:}} \; \mu_{\ell }=\frac{\sum\limits_{n  = 1}^{N_{\rm elite}} \sum\limits_i
\varphi _i^{\ell ,n} f_i^{(n) }} {\sum\limits_{n  = 1}^{N_{\rm elite}}  \sum\limits_i \varphi _i^{\ell ,n}},  \label{GMM-M1} \\
&\quad\quad\quad  \vartheta_{\ell }=\frac{\sum\limits_{n  = 1}^{N_{\rm elite}} \sum\limits_i \varphi _i^{\ell ,n} \left(f_i^{(n)} -\mu_{\ell }\right)^2 } {\sum\limits_{n  = 1}^{N_{\rm elite}} \sum\limits_i \varphi _i^{\ell ,n}},\\
&\quad\quad\quad {\varpi _{\ell } }=\frac{\sum\limits_{n  = 1}^{N_{\rm elite}} \sum\limits_i \varphi _i^{\ell ,n} } {N_{\rm elite} {\rm dim(\mathbf{f})}}. \label{GMM-M3}
\end{align}
In the GMM, there are different candidate models with different numbers of mixture components $\varsigma$, which needs to be delicately selected.  Since Bayesian information criterion (BIC) is widely adopted to evaluate the candidate model~\cite{BIC2012}, BIC can be a means to adaptively determine $\varsigma$, which is calculated as
\begin{align}\label{BIC}
{\rm BIC}\left( \varsigma\right)=3\varsigma \ln\left(N_{\rm elite}\right)-2\ln\left(\mathcal{L}\left(\mathbf{f}|{\bm \varpi},{\bm \mu},{\bm \vartheta}\right)\right),
\end{align}
where $\mathcal{L}\left(\mathbf{f}|{\bm \varpi},{\bm \mu},{\bm \vartheta}\right)$ is the  likelihood function for a candidate GMM.  Thus the adaptive GMM is obtained, namely the best candidate model with the minimum BIC value. As such, we solve the problem \eqref{Opt_problem} by using the proposed \textbf{Algorithm 1}, to obtain the optimal solution $\{\mathbf{f}^*,\mathbf{B}^*\}$. In \textbf{Algorithm 1}, when there exists frequency overlap between adjacent subchannels, the overlapped frequency band is allocated to the subchannel with larger RSS at its center frequency.
{  \begin{algorithm}[htp] \label{algorithmic1}
  Initialize Gaussian distributions' parameters {\small{$\left\{\mathcal{N}\left(\mu_n=f_{\rm co}+(n-\frac{1}{2})\frac{f^{\rm upper}-f_{\rm co}}{N_{\rm c}}, \delta_n^2=\frac{\left(f^{\rm upper}-f_{\rm co}\right)^2}{4 {N_{\rm c}}^2}\right)\right\}_{n=1}^{N_{\rm c}}$}} with the total number of subchannels $I$; the number of samples $N_{\rm c}$, and the number of elite samples $N_{\rm elite}$  that have better objective values than other samples; the iteration index $t=0$\\
\While{$t < t^{\rm max}$}{i) Generate $\mathbf{S}_i $ samples according to the GMM $\mathbf{S}_i\sim p\left(f|{\bm \varpi},{\bm \mu},{\bm \vartheta}\right)$, then, given $\mathbf{S}_i$, obtain each sample's corresponding subchannel bandwidths  according to one-dimensional search under the constraints $\mathrm{C1}$--$\mathrm{C5}$;\\
ii) { Select $\mathbf{\widehat{S}}_i  $ elite samples from $\mathbf{S}_i$ that have better objective values of \eqref{Opt_problem} than other samples}. Based on elite samples of the current iteration and the   best elite sample with the largest reward at the prior iteration, estimate the GMM parameters $\{{\bm \varpi},{\bm \mu},{\bm \vartheta}\}$ via EM algorithm given by \eqref{GMM-M1}--\eqref{GMM-M3} and its adaptive one with BIC given by \eqref{BIC}, then, the estimated parameters are updated in a smoothing manner:
\\
iii) $t=t+1$;}
 {Optimal $\{\mathbf{f}^*,\mathbf{B}^*\}$ with the largest award is obtained}
 \caption{\small{Spatial-spectral Cell-free Subchannel Allocation}}
\end{algorithm}
}

\section{AP Clustering and Subchannel Allocation in the Large-scale Cell-free Networks}
\begin{figure}
     \centering
    \subfigure[High-dimensional space without AP clustering.]{
         \centering
         \includegraphics[width=2.7 in,height=1.8 in]{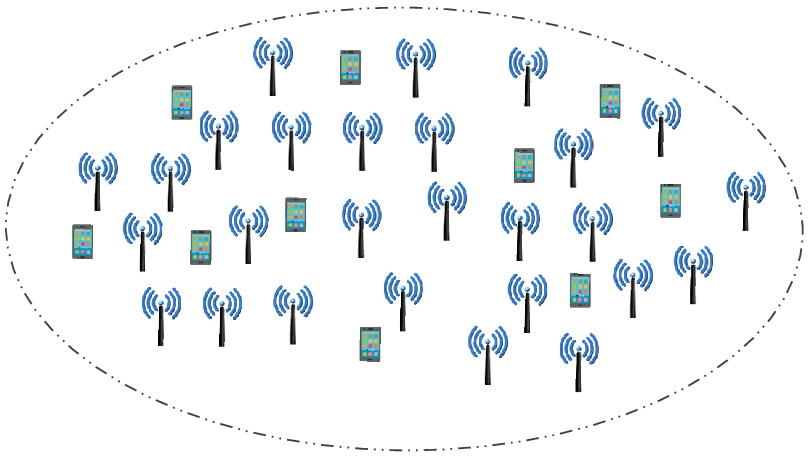}
      \label{figa1_k}
     }
     \subfigure[AP clustering with dimensionality reduction.]{
         \centering
         \includegraphics[width=2.7 in,height=1.8 in]{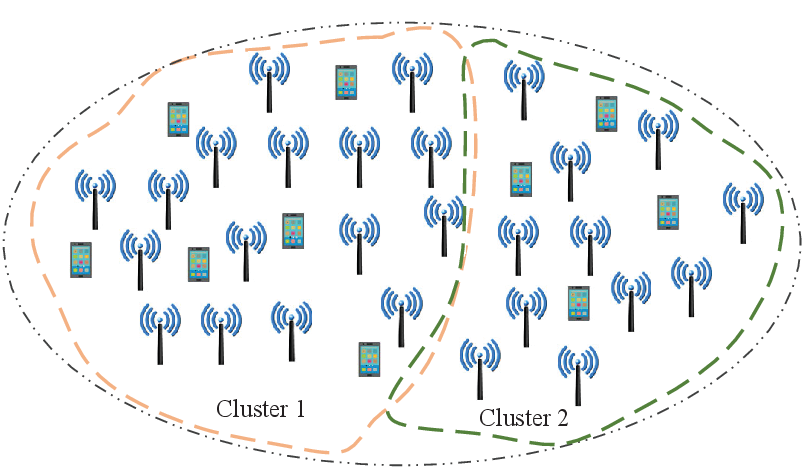}
      \label{figb1_k}
    }
 \caption{The large-scale cell-free networks without/with AP clustering.}
 \label{Fig_split_antenna}
\end{figure}

In the prior Section~\ref{SA-CF}, subchannel allocation  is designed by considering a large-scale spatial-spectral cell-free massive MIMO system without addressing the initial access and realistic networking architecture consisting of many cell-free subnetworks (AP clusters). 6G cell-free networking architecture may evolve in terms of the different cooperation levels and AP clustering plays a crucial role in practical deployment~\cite{JTAO2023}.  In the large-scale cell-free mmWave/THz downlink networks with dense distributed APs, AP clustering with inter-CPU coordination is indispensable for scalability, i.e., an AP cluster consisting of a subset of APs coordinately serves multiple UEs and each cluster is controlled by a single CPU~\cite{Ranjbar2022}.  In addition, clustering helps to reduce the signal processing complexity for large-scale case, since each cluster's cell-free massive MIMO processing is in a lower-dimensional space as shown in Fig.~\ref{Fig_split_antenna}. User-centric clustering in the lower-frequencies has been studied in the literature~\cite{Yanlin2019,Junyuan2023}, for instance, \cite{Yanlin2019} provides the coverage distance criterion for clustering,  and~\cite{Junyuan2023} seeks to maximize the number of clusters under the quality of service (QoS) constraint,  however, these designs make that cell-free massive MIMO baseband processing is sensitive to the pre-defined QoS threshold and an unappropriate QoS threshold may degrade the massive MIMO. Network-centric clustering is investigated in~\cite{Michal2021}, where the radius for a cluster of base stations (BSs) is calculated as the minimal maximal distance between BSs in the same cluster and each user is associated with the BS by following the best channel rule, however, such a design forms fixed AP clusters and may lead to the overloading issue in some clusters with the dense UEs, which is not scalable. The AP clustering scheme in~\cite{Zaher2024} is based on the RSS and its hyperparameter is the maximum AP cluster size, which has to be delicately tuned by the radio network controller.
Partitioning and hierarchical approaches are two widely-adopted types of clustering, the former selects the fixed number of clusters, and the latter makes hierarchical decomposition involving the  merge process~\cite{Xiaowei1996}. Therefore, we first investigate the AP clustering algorithms under these two types.

\subsection{$\mathcal{K}$-means Clustering}\label{KNN}
 $\mathcal{K}$-means algorithm is the typical partitioning clustering design, where the hyperparameter $\mathcal{K}$ is the pre-defined number of clusters.  One benefit of  $\mathcal{K}$-means based AP clustering is that the total wireless fronthaul cost  in a cluster is minimal if the gravity center of the cluster is the CPU's location. For millimeter wave or terahertz cell-free networks with narrow beams in which small-scale fading is negligible, $\mathcal{K}$-means based AP clustering constructs a voronoi diagram, namely the shape of an AP cluster $\mathcal{C}_z$ ($z=1,\cdots,\mathcal{K}$) is the voronoi cell. To determine the association between each UE and the AP cluster, we adopt a fast initial access approach with one-shot discovery based on the leaky-wave antenna's radiation property. Specifically, the frequency $f_{mk}^{\rm max}$ for achieving the maximum level of radiation of a link from AP $m$ to UE $k$ is given by~\cite{Yasaman_2020}
 \begin{align}\label{radia}
 f_{mk}^{\rm max}=\frac{f_{\rm co}}{\sin\left(\theta_{mk}\right)}.
 \end{align}
 Thus the corresponding RSS $\gamma_{mk}^{\rm RSS}$ of a link from AP $m$ to UE $k$ at the frequency $f_{mk}^{\rm max}$ can be calculated.
 \begin{align}\label{RSS12}
 \gamma_{mk}^{\rm RSS}=q_k G\left( {f_{mk}^{\rm max},\theta _{mk} } \right) |h_{mk}|^2.
 \end{align}
 According to the RSS-based rule, each UE is associated with the AP that provides the maximum RSS, i.e., the desired AP $m^*=\mathop {\arg \max }\limits_m \{\gamma_{mk}^{\rm RSS}\}$, thus each UE is served by the AP cluster including its desired AP. It is noted that the distributed massive MIMO links between the UEs and the APs in a cluster can also be rapidly discovered with the help of the leaky-wave antenna's radiation property~\cite{LZ20222023}.
\subsection{Hierarchical Clustering}\label{H-cluster}
The partitioning clustering approach such as $\mathcal{K}$-means fixes the number of clusters, which is not scalable. Moreover, the initial access is carried out after fulfilling the AP clustering, which means that the diversity of access links is not well exploited during the AP clustering process. To cope with these issues, we propose a hierarchical clustering approach consisting of the following three steps:
  \begin{description}
    \item[\textbf{Step 1:}]~~The initial clusters are formed by leveraging affinity propagation~\cite{Brendan2007}, which is an efficient approach to adaptively find the network-centric clusters without the need to initialize the number of AP clusters. During the message-passing procedure of affinity propagation, the similarity $s(m,m^{'})$ between AP $m$ and its exemplar AP $m^{'}$ is evaluated as a negative Euclidean distance between them, and the ``preferences'' is the median of the similarities under the assumption that all the APs have an equal chance of being exemplars. Thus the network-centric clusters $\{\mathcal{C}_z\}$ are obtained, based on $\{\mathcal{C}_z\}$, each UE is associated with the AP following the RSS-based rule of subsection~\ref{KNN}.
    \item[\textbf{Step 2:}]~~Based on the AP clusters $\{C_{z }\}$, some AP clusters denoted by $\{\mathcal{C}_z^{\rm void}\}$ may not serve any UEs, which should be merged with the rest of AP clusters denoted by $\{\mathcal{C}_{z^{'} }\}$ that serve UEs. Given an AP cluster $\{\mathcal{C}_z^{\rm void}\}$,  it is chosen to be merged with an AP cluster in  $\{\mathcal{C}_{z^{'} }\}$ such that the spectral efficiency per AP in this cluster is maximized~\cite{Wagner2012}, which is written as
        \begin{align}\label{SE_AP}
\hspace{-0.85 cm}C_{z^{'} }^{*}  = \mathop {\arg \max }\limits_{C_{z^{'}}^{\rm merge} } \frac{1}{{{\rm card}(C_{z^{'}}^{\rm merge} )}}\sum\limits_{k \in C_{z^{'}}^{\rm merge} } {\log _2 \left( {1 + \gamma _k \left( f \right)} \right)},
\end{align}
where $C_{z^{'}}^{\rm merge} $ denotes the merged cluster and ${\rm card}(C_{z^{'}}^{\rm merge} )$ is the number of APs in this cluster. The merge criterion given by \eqref{SE_AP} ensures that APs can benefit from the coordination, and can be instinctively understood by the fact that an AP belongs to a cluster since it brings in more diversity gains than the case of being located in other clusters. Each cluster in $\{C_{z^{'} }^*\}$ serves a number of UEs. In this way, the number of updated AP clusters $\{C_{z^{'} }^*\}$ shall be not greater than the $\{\mathcal{C}_{z^{'} }\}$.
    \item[\textbf{Step 3:}]~~In light of the AP clusters $\{C_{z^{'} }^*\}$, two AP clusters are merged if  the spectral efficiency per AP is maximally improved, which is given by
    \begin{align}\label{SE_AP11}
\hspace{-0.85 cm} C_{z^{''} }^{*}  = \mathop {\arg \max }\limits_{C_{z^{'}}^{*\rm merge} } \frac{1}{{{\rm card}(C_{z^{'}}^{*\rm merge} )}}\sum\limits_{k \in C_{z^{'}}^{*\rm merge} } {\log _2 \left( {1 + \gamma _k \left( f \right)} \right)},
\end{align}
where $C_{z^{'}}^{*\rm merge} $  denotes the merged cluster and ${{\rm card}(C_{z^{'}}^{*\rm merge} )}$ is its number of APs. As such, the number of ultimate AP clusters $\{C_{z^{''} }^*\}$ is further reduced to be not greater than the $\{C_{z^{'} }^*\}$.
  \end{description}

Differing from the partitioning clustering approach in subsection~\ref{KNN},  the proposed hierarchical clustering has addressed the effects of both AP positions and access links on the AP clustering, namely a joint initial access and cluster formation approach.

 After fulfilling the initial access and cluster formation, each AP cluster's baseband processing is carried out by one single CPU, unfortunately, interference management between AP clusters may still need to be addressed. Thanks to the very large bandwidths at sub-THz bands, frequency reuse scheme can be adopted to avoid the co-channel interference. However, orthogonally allocating the frequency bandwidths to the AP clusters is problematic. By extending the line of work in Section~\ref{SA-CF}, the subchannel allocation problem after forming the AP clusters is written as
\begin{align}\label{Opt_problem-cluster}
&\mathop {\max }\limits_{{\bf{B,f}}} R_{\rm total} = \sum\limits_i {B_{i} } \sum\limits_{{k,i}\in \{C_{z^{''} }^*\}} \log_2\left(1+\gamma _{k}\left(f_i\right)\right)\\
&\mathrm{s.t.} ~\mathrm{C1},~\mathrm{C2},~\mathrm{C3},~\mathrm{C4},~\mathrm{C5},\nonumber\\
&\mathrm{C6:}~\frac{\sum\limits_{{k,i}\in C_{z^{''} }^*} B_{i} \log_2\left(1+\gamma _{k}\left(f_i\right)\right)}{{\rm card}(\{k\in C_{z^{''} }^*\} )}\geq \bar{R}_{\rm th},~~\forall C_{z^{''} }^*, \nonumber
\end{align}
where constraint $\mathrm{C6}$ makes sure that each AP cluster achieves the minimum average rate per UE such that the uniformly good services in cell-free networks are not compromised. Compared with the prior problem \eqref{Opt_problem}, problem \eqref{Opt_problem-cluster} has to bring in additional constraint $\mathrm{C6}$ for inter-cluster coordination. Problem \eqref{Opt_problem-cluster} is highly combinatorial, hence a suboptimal approach is proposed by updating \textbf{Algorithm 1} in Section~\ref{SA-CF}. Specifically, in step 3 of \textbf{Algorithm 1},  given a sample of the subchannel center frequencies, sample's corresponding subchannel bandwidths in each cluster can be obtained according to one-dimensional search under the constraints $\mathrm{C1}$--$\mathrm{C5}$, by sorting the AP clusters in descending order based on the reward (objective in \eqref{Opt_problem-cluster}) of a candidate subchannel in the sample, the candidate subchannel is solely allocated to an AP cluster by following the two cases:
\begin{itemize}
  \item Case 1: The constraint $\mathrm{C6}$ should be firstly feasible, i.e.,  AP clusters that have not met $\mathrm{C6}$ have the higher priority to occupy the candidate subchannel, among these clusters, the candidate subchannel is preferred to be dedicated to an AP cluster with the largest award (maximum transmission rate in such a candidate subchannel);
  \item Case 2: After meeting the feasibility of problem \eqref{Opt_problem-cluster} (namely all the clusters reach the minimum QoS level), a candidate subchannel  is dedicated to the AP cluster with the largest award.
\end{itemize}
The other steps of \textbf{Algorithm 1} stay the same. Thus the effects of AP clustering on the subchannel allocation  and interference management between AP clusters are addressed.

\section{Numerical Results}
This section presents numerical results to demonstrate the efficiency of the proposed solutions. In the simulations, APs and UEs are
independently and uniformly distributed in an area of $200$m$\times 200$m, the absolute antenna elevation difference between the AP
and UE follows the uniform distribution with the interval $[7,10]$m, and the propagation angle from the $m$-th AP to the $k$-th UE is uniformly distributed, i.e., $\theta_{km} \in U(0, \frac{\pi}{2})$. Free-space path loss model is employed~\cite{LZ2022}, and the leaky-wave antenna's attenuation coefficient is $\varrho=130$rad/m, aperture length is $L=0.15$m, and $\xi=1$. Two sub-THz frequency bands are considered, namely $f \in \left[100,200\right]$GHz with the cutoff frequency $f_{\rm co}=100$GHz, and $f \in \left[200,300\right]$GHz with the cutoff frequency $f_{\rm co}=200$GHz. The other basic parameters are shown in Table I.

{\begin{flushleft}
\begin{table}[h]\label{table1}
\centering
\caption{Simulation parameters}
\setlength\tabcolsep{3pt} \begin{tabular}{|l|c|}
  \hline
  Total downlink transmit power & $P_{\rm total}=\bar{q}*B_{\rm total}=2$W\\ \hline
  Equal transmit power allocation & $q_{k}=\bar{q}$,~$\forall k$ \\ \hline
  Noise's PSD & $\delta ^2=-168$dBm/Hz\\ \hline
 Threshold for the minimum receive power& $\bar{\gamma}_{\rm th}=-174$dm/Hz\\ \hline
 RSS gap in the frequencies of a subchannel  & $\varepsilon=0.5$dB \\
  \hline
\end{tabular}
\end{table}
\end{flushleft} }

\subsection{Subchannel Allocation without AP Clustering}\label{simu_A}
In this subsection, we evaluate the proposed subchannel allocation solution in a large-scale cell-free system as illustrated in Section III. The traditional cross-entropy design~\cite{Zesheng2024} and equal bandwidth allocation~\cite{Chia-Yi_TFS} are the benchmarks.

\begin{figure}
     \centering
    \subfigure[MRT precoding.]{
         \centering
         \includegraphics[width=3.0 in,height=2.3 in]{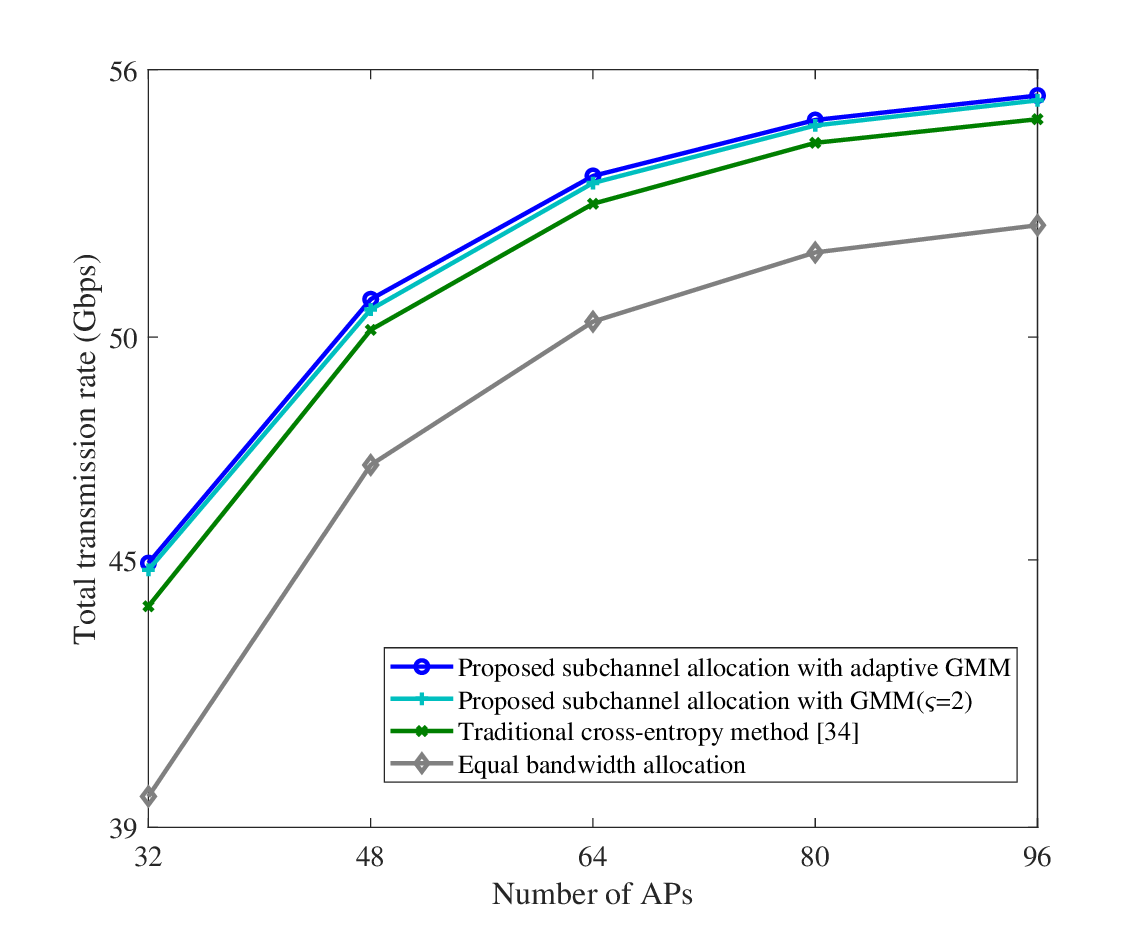}
      \label{figa1_k}
     }
     \subfigure[ZF precoding.]{
         \centering
         \includegraphics[width=3.0 in,height=2.3 in]{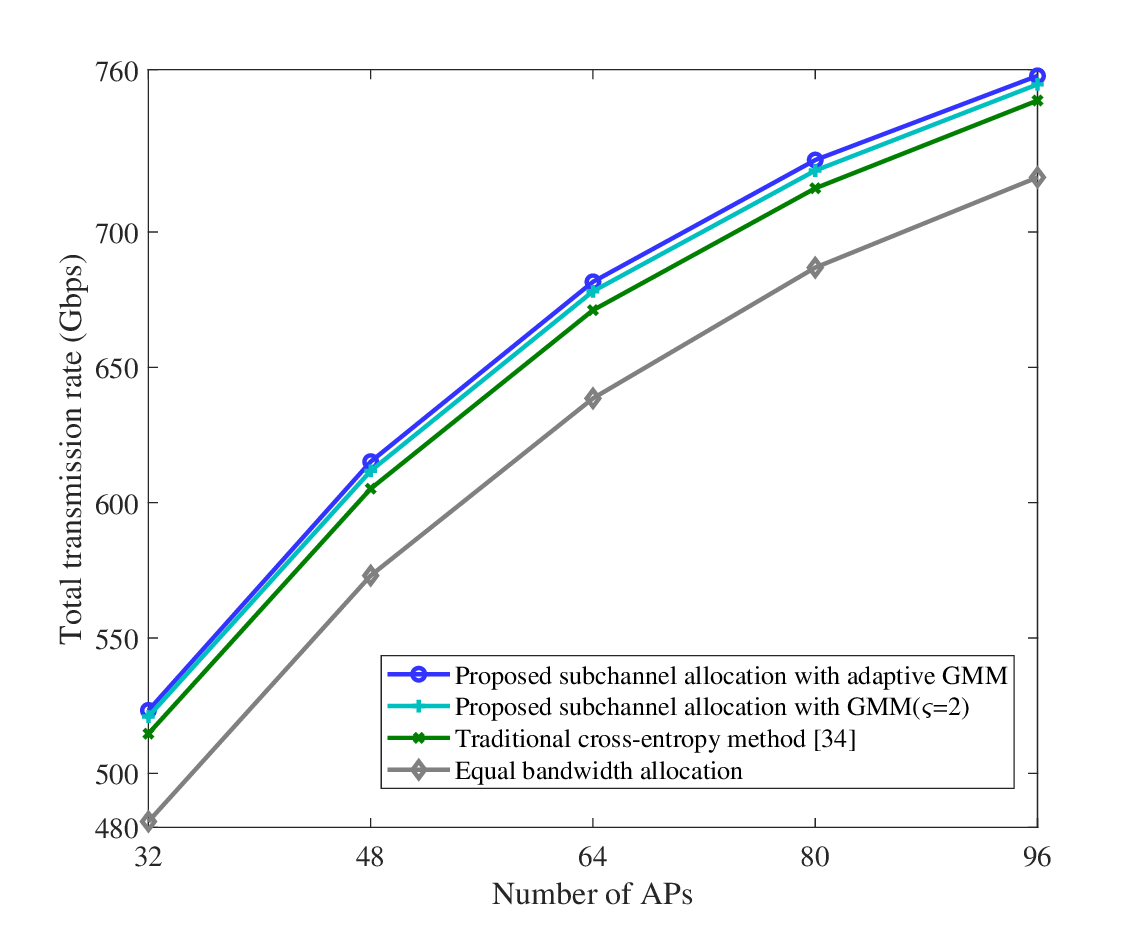}
      \label{figb1_k}
    }
 \caption{The total transmission rate versus number of APs under different precoding methods with $f_{\rm co}=100$GHz, $B_{\rm total}=10$GHz and ten UEs.}
 \label{Fig_num_antenna}
\end{figure}

Fig.~\ref{Fig_num_antenna} shows the total transmission rate versus number of APs under MRT and ZF precoding methods when the frequencies are located in the range $\left[100,200\right]$GHz with the cutoff frequency $f_{\rm co}=100$GHz. It is known that deploying more APs creates more antenna gains, which improves the spectral efficiency. The proposed subchannel allocation solution with adaptive GMM achieves the best performance, compared with the baseline solutions. MRT precoding method has the lowest complexity and achieves impressive total transmission rate in the cell-free system, however, its performance is still even less than the ZF due to the inter-UE interference. As revealed in Fig.~\ref{figa1_k}, optimizing subchannel allocation can greatly improve the total transmission rate when adopting MRT precoding in the cell-free system.

\begin{figure}
     \centering
    \subfigure[MRT precoding.]{
         \centering
         \includegraphics[width=3.0 in,height=2.3 in]{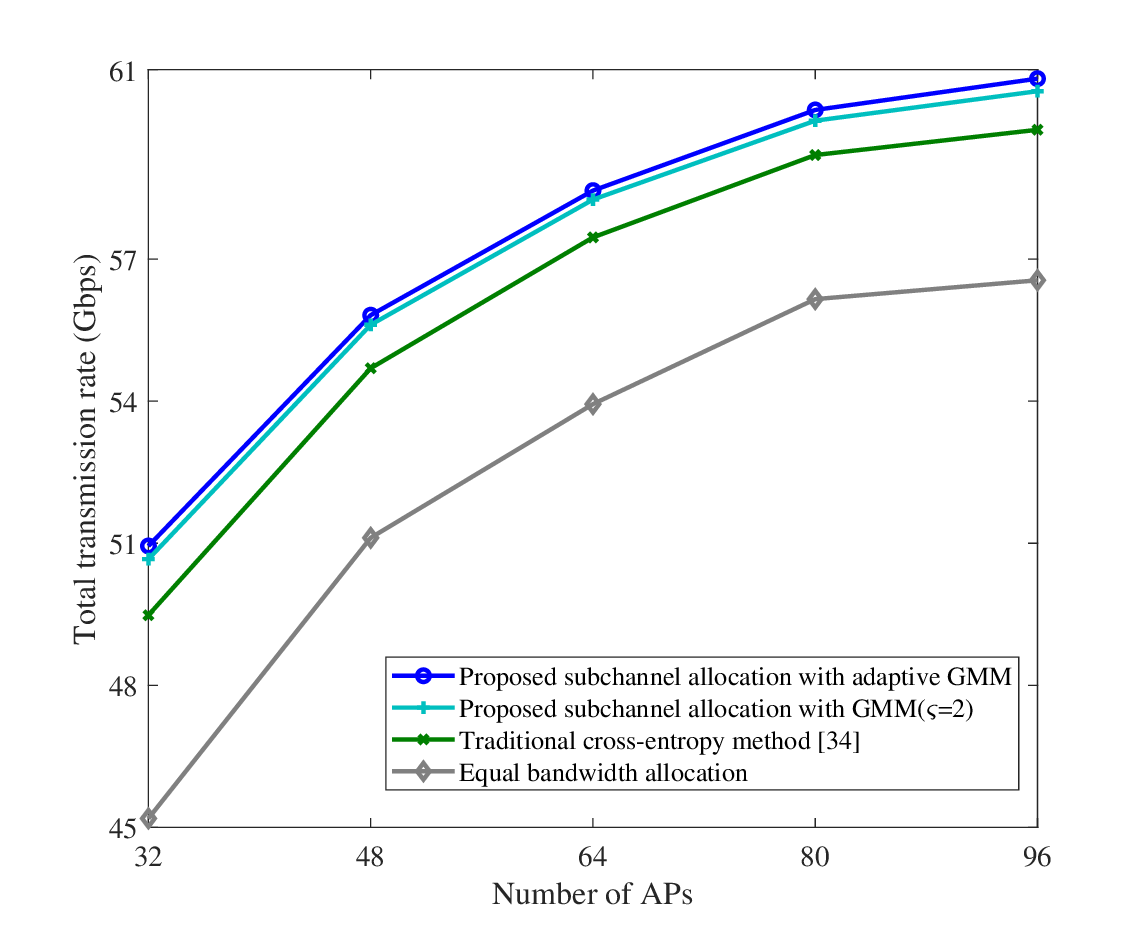}
      \label{figa2_k}
     }
     \subfigure[ZF precoding.]{
         \centering
         \includegraphics[width=3.0 in,height=2.3 in]{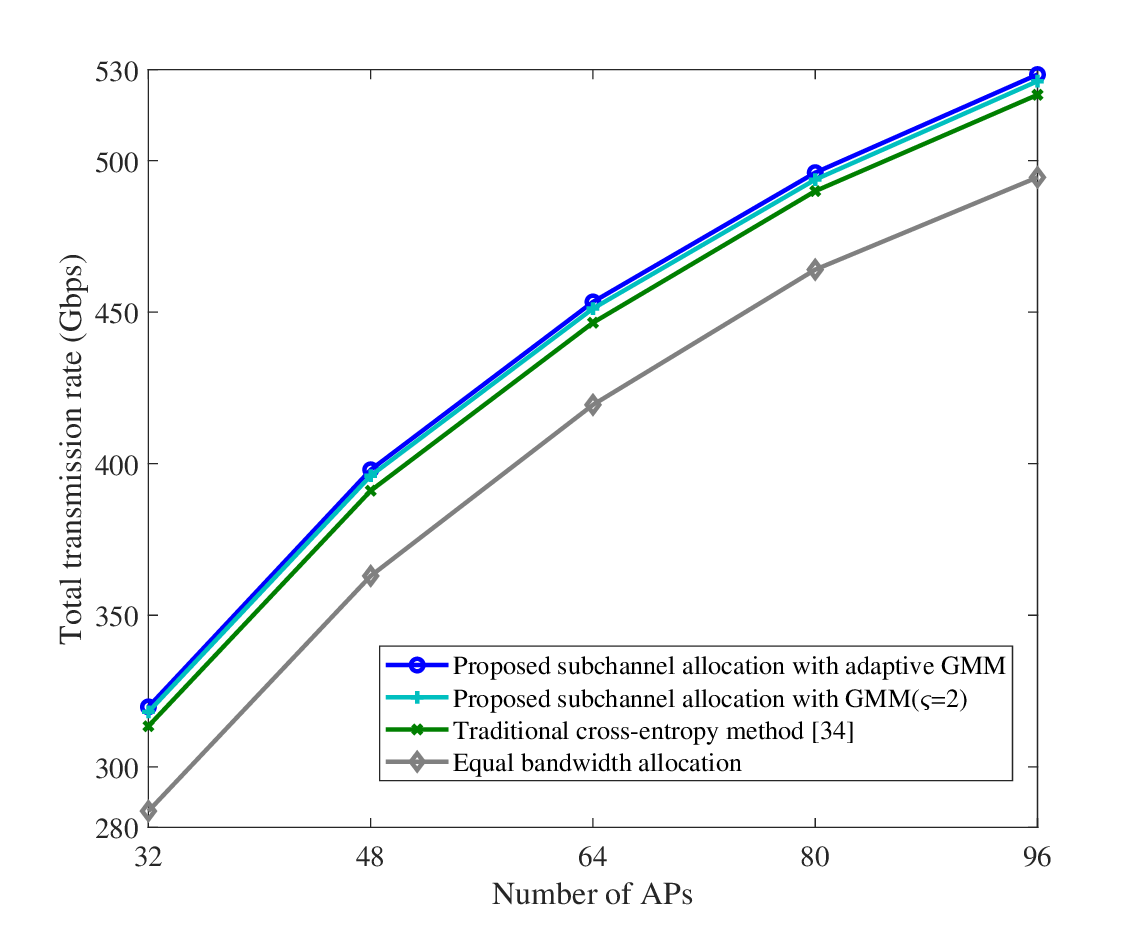}
      \label{figb2_k}
    }
 \caption{The total transmission rate versus number of APs under different precoding methods with $f_{\rm co}=200$GHz, $B_{\rm total}=10$GHz and ten UEs.}
 \label{Fig_num_fco}
\end{figure}

Fig.~\ref{Fig_num_fco} shows the total transmission rate versus number of APs under MRT and ZF precoding methods when the frequencies are located in the range $\left[200,300\right]$GHz with the cutoff frequency $f_{\rm co}=200$GHz. It has the similar performance trend as seen in Fig.~\ref{Fig_num_antenna}, and there exists an interesting phenomenon that total transmission rate achieved by the MRT precoding at higher frequency bands is better than the case of using lower frequency bands in Fig.~\ref{figa1_k}, which can be explained by the fact that the level of the inter-UE interference is less in the higher-frequencies. By comparing Fig.~\ref{figb2_k} with Fig.~\ref{figb1_k}, it is obviously seen that when ZF precoding is applied to cancel the inter-UE interference,  the higher path loss in the higher-frequencies is a key factor to degrade the transmission rate.

\begin{figure}
     \centering
    \subfigure[Cutoff frequency $f_{\rm co}=100$GHz.]{
         \centering
         \includegraphics[width=3.0 in,height=2.3 in]{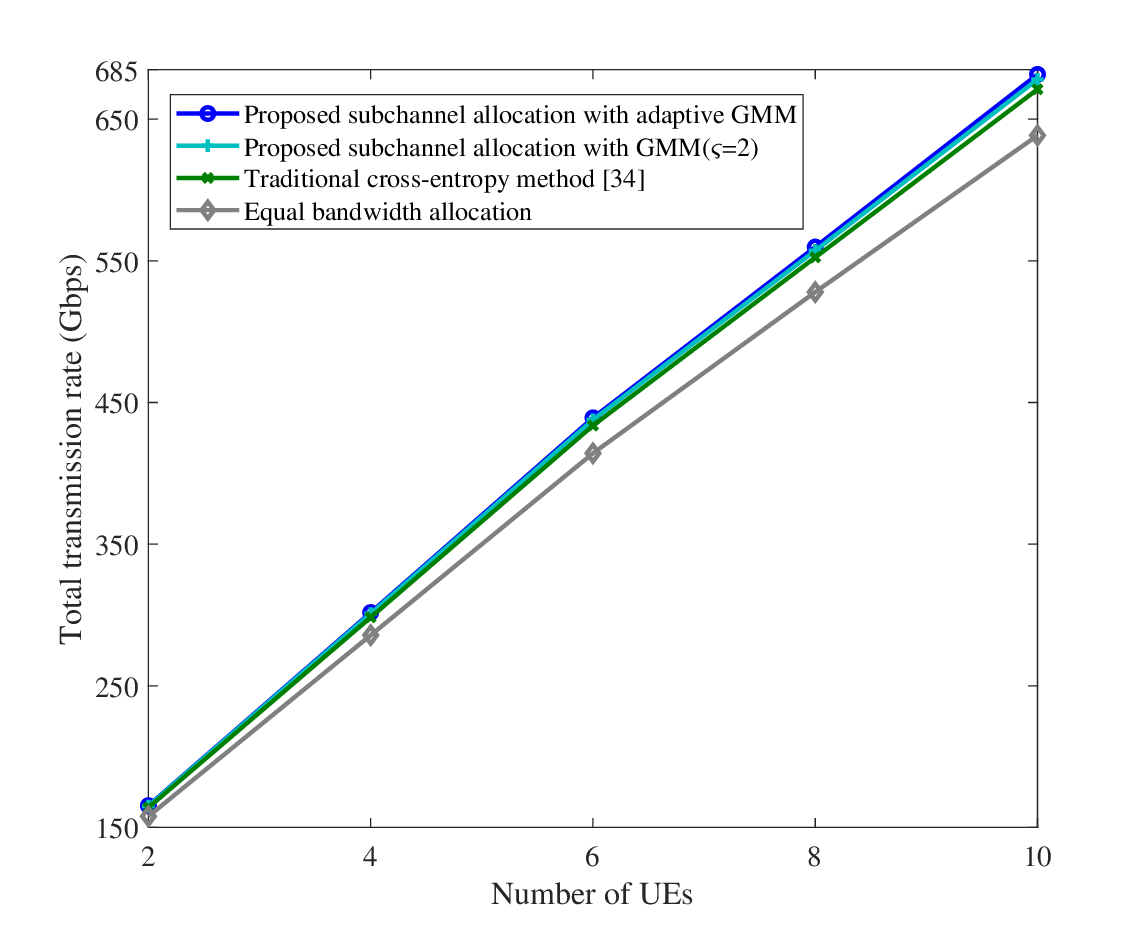}
      \label{figa3_k}
     }
     \subfigure[Cutoff frequency $f_{\rm co}=200$GHz.]{
         \centering
         \includegraphics[width=3.0 in,height=2.3 in]{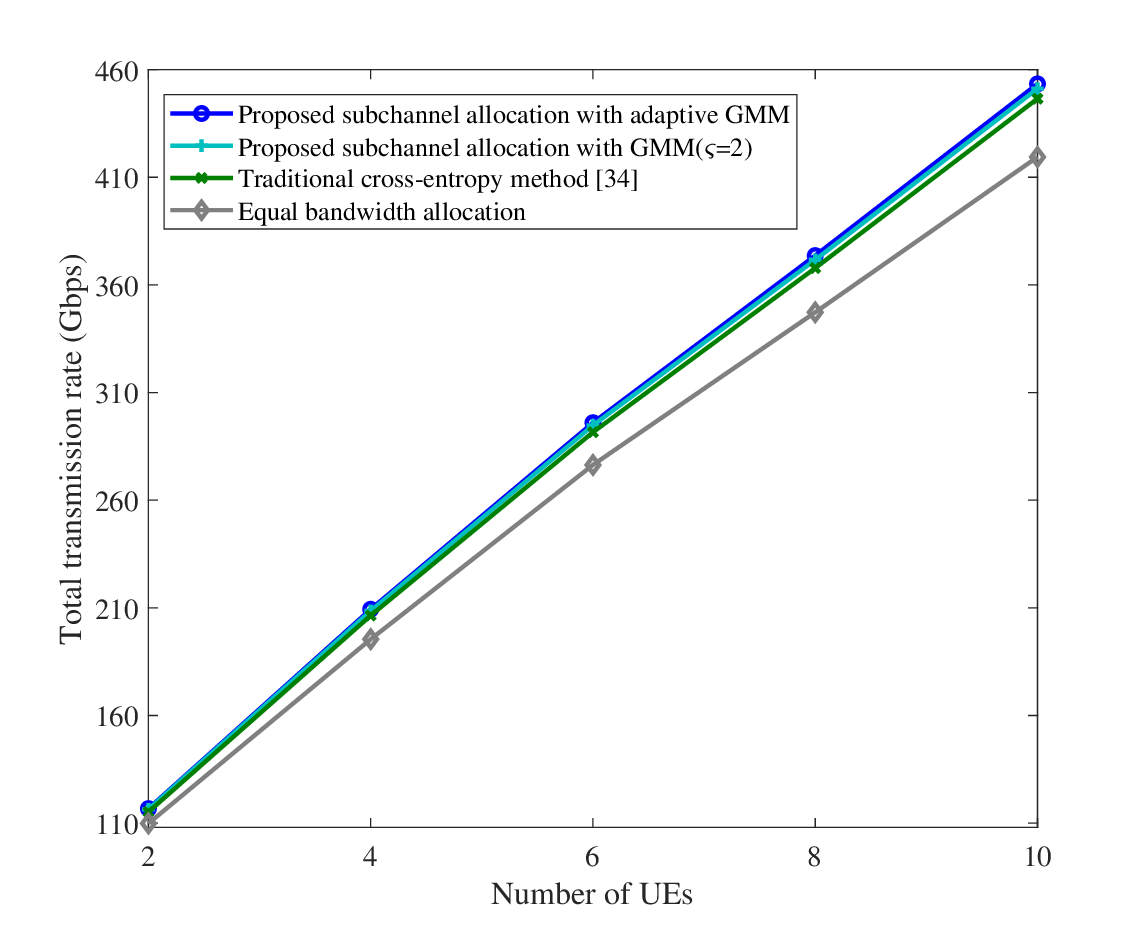}
      \label{figb3_k}
    }
 \caption{The total transmission rate versus number of UEs under different sub-THz frequency bands with $M=64$ and $B_{\rm total}=10$GHz.}
 \label{Fig_num_UE200}
\end{figure}

Fig.~\ref{Fig_num_UE200} shows the total transmission rate versus number of UEs under different sub-THz frequency bands when adopting the ZF precoding. The total transmission rate increases with the number of UEs due to the multiuser diversity, namely larger multiplexing gains are obtained. When serving more UEs, optimizing subchannel allocation helps to improve the performance, and its advantage over the equal bandwidth allocation grows large. Again, we see that the higher path loss in the higher frequencies results in lower transmission rate under the ZF precoding.

\subsection{AP Clustering with Subchannel Allocation}
In this subsection, we evaluate the proposed Ap clustering with the subchannel allocation solution in a large-scale cell-free network as illustrated in Section IV, where the optimal subchannel allocation is carried out by leveraging the adaptive GMM as confirmed in the subsection~\ref{simu_A}. The $\mathcal{K}$-means methods with different pre-set numbers of clusters are the benchmarks. Given a specific clustering approach, the proposed optimal subchannel allocation in  Section IV is in comparison with the equal bandwidth allocation~\cite{Chia-Yi_TFS}. The ZF precoding process is independently executed in each cluster.

\begin{figure}
     \centering
    \subfigure[Cutoff frequency $f_{\rm co}=100$GHz.]{
         \centering
         \includegraphics[width=3.0 in,height=2.3 in]{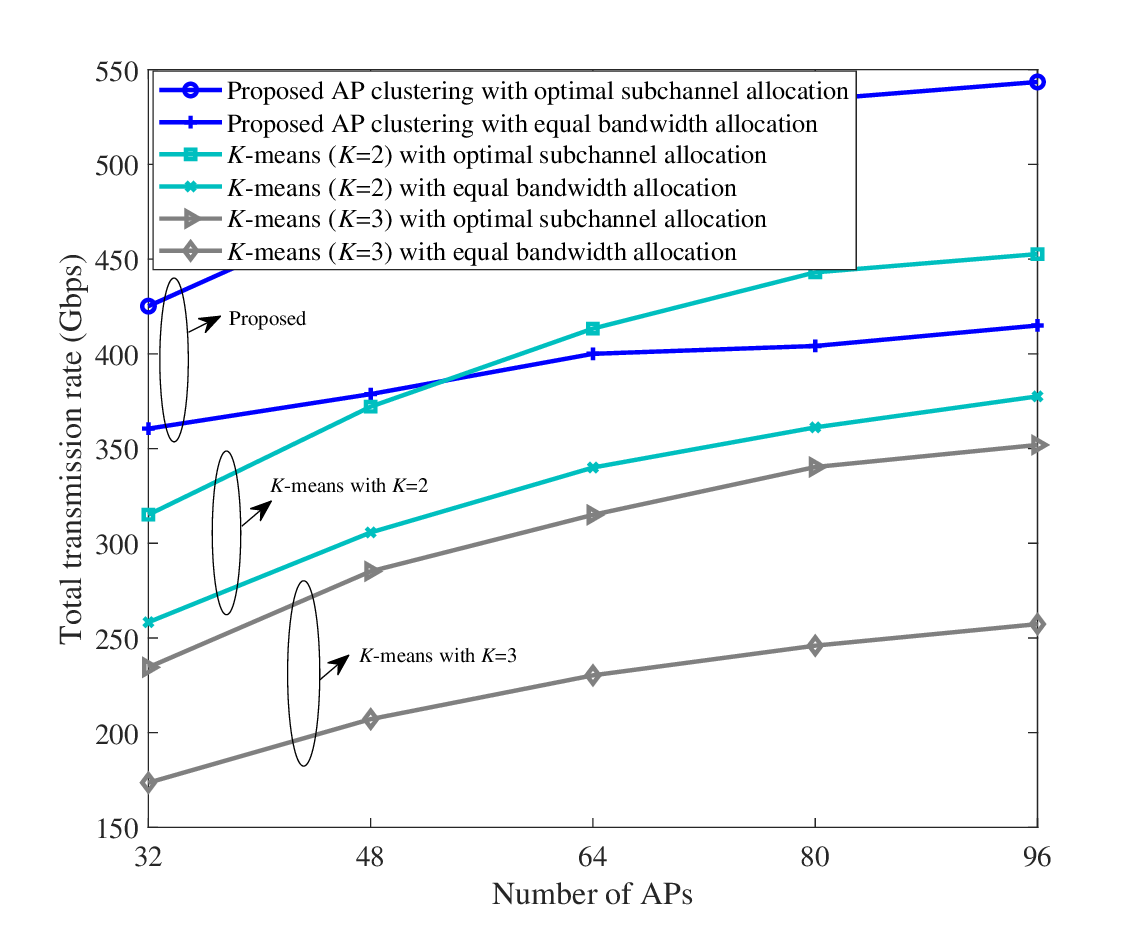}
      \label{figa3_APc}
     }
     \subfigure[Cutoff frequency $f_{\rm co}=200$GHz.]{
         \centering
         \includegraphics[width=3.0 in,height=2.3 in]{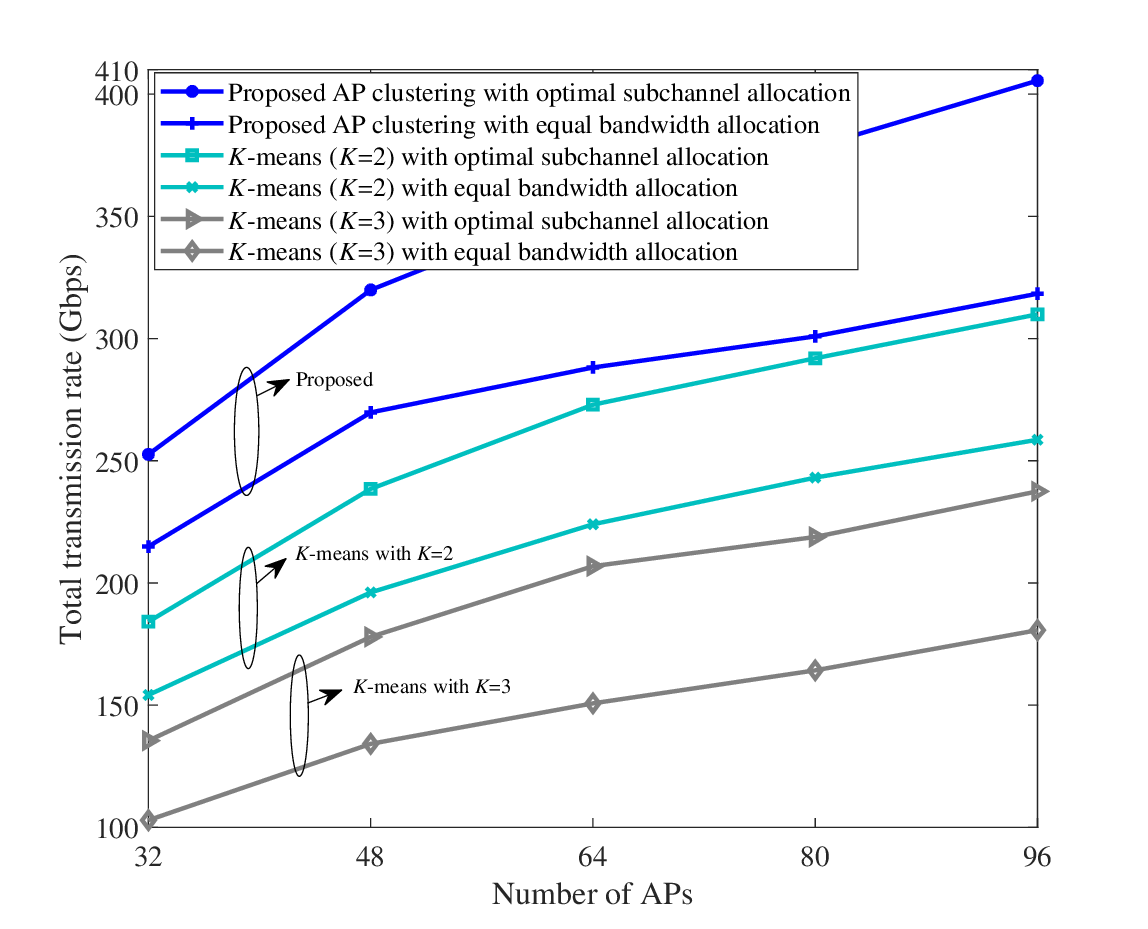}
      \label{figb3_APc}
    }
 \caption{The total transmission rate versus number of APs under different sub-THz frequency bands with $B_{\rm total}=10$GHz and ten UEs.}
 \label{Fig_num_APcluster200}
\end{figure}

Fig.~\ref{Fig_num_APcluster200} shows the  total transmission rate versus number of APs under different sub-THz frequency bands. The proposed AP clustering with optimal subchannel allocation achieves higher transmission rate than other solutions, and the transmission rate increases when deploying more APs. Compared to the $\mathcal{K}$-means clustering, the proposed AP clustering can significantly enhance the data rate. Given the specific clustering approach, the proposed subchannel allocation in Section IV can well manage the interference between clusters and has better performance than the equal bandwidth allocation. Due to the higher path loss in the higher-frequencies, the transmission rate decreases and more APs are required to reach the similar level of the transmission rate compared with the counterparts in the lower-frequencies. More importantly, compared with Fig. \ref{figb1_k} and Fig. \ref{figb2_k} in which all the APs simultaneously serve all the UEs in a holistic and unpractical manner with the largest link costs, the proposed clustering with optimal subchannel allocation solution reduces the scale of massive MIMO links and thus the link costs at each cluster without suffering much network performance loss. In addition, it is indicated from Fig.~\ref{figb3_APc} that  the proposed AP clustering plays a dominant role in the performance enhancement when the path losses of the cell-free links are very large and have to be astutely addressed at higher THz bands.

\begin{figure}
     \centering
    \subfigure[Cutoff frequency $f_{\rm co}=100$GHz.]{
         \centering
         \includegraphics[width=3.0 in,height=2.3 in]{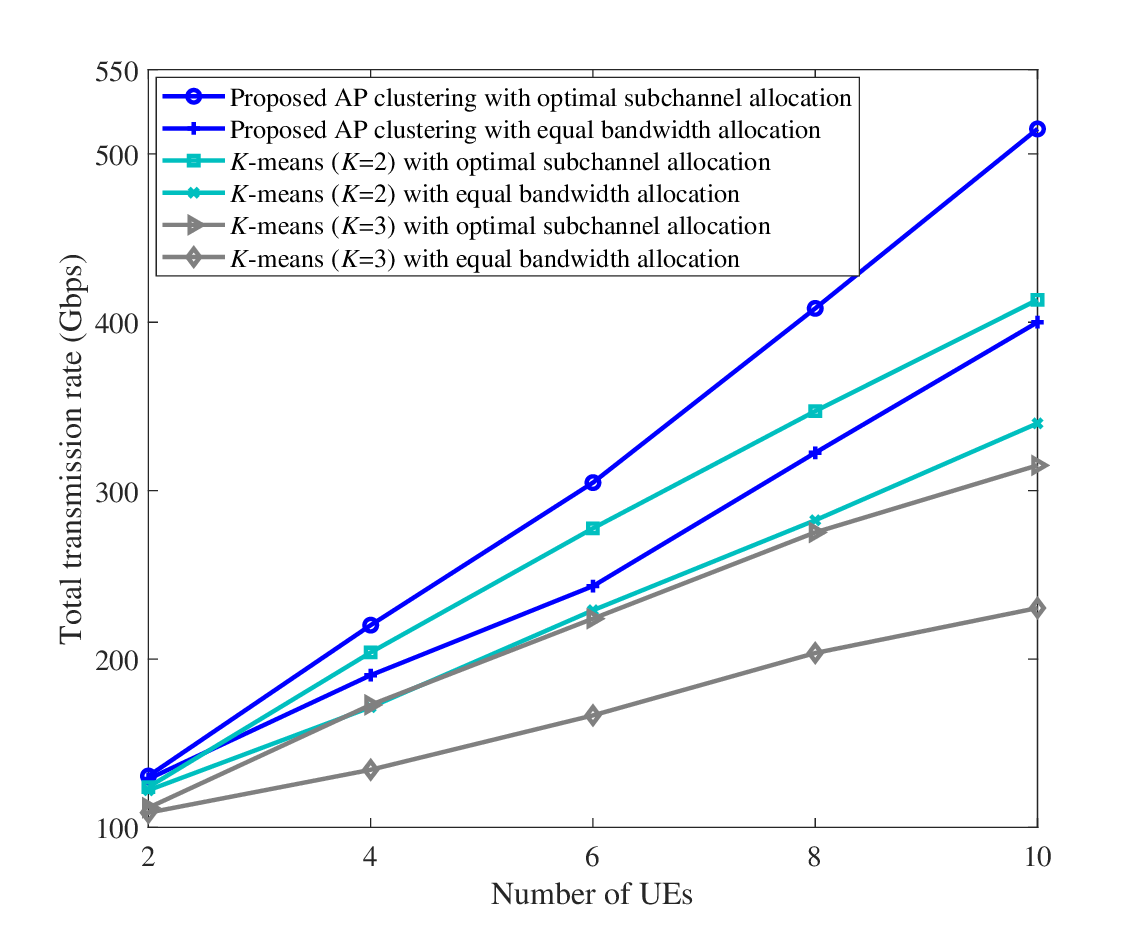}
      \label{figa3_UEC}
     }
     \subfigure[Cutoff frequency $f_{\rm co}=200$GHz.]{
         \centering
         \includegraphics[width=3.0 in,height=2.3 in]{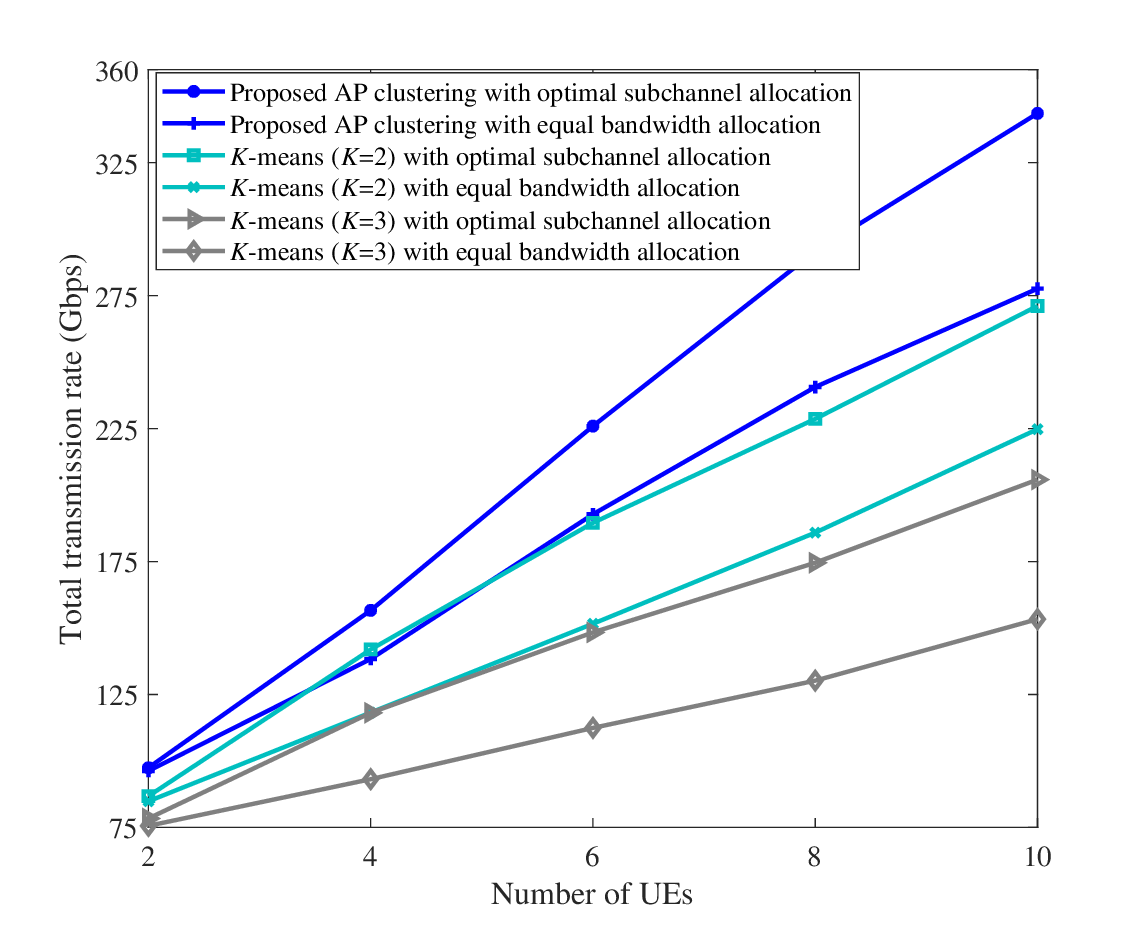}
      \label{figb3_kUEC}
    }
 \caption{The total transmission rate versus number of UEs under different sub-THz frequency bands with $M=64$ and $B_{\rm total}=10$GHz.}
 \label{Fig_num_UEcluster200}
\end{figure}

Fig.~\ref{Fig_num_UEcluster200} shows the  total transmission rate versus number of UEs under different sub-THz frequency bands. The proposed AP clustering with optimal subchannel allocation performs the best transmission rate at the sub-THz bands. Both AP clustering and subchannel allocation can have a big impact on the total transmission rate and such an impact becomes more significant as the number of UEs increases, moreover, the proposed AP clustering with equal bandwidth allocation could achieve better performance than the case of using $\mathcal{K}$-means clustering with optimal subchannel allocation as shown in Fig.~\ref{figa3_UEC}. For $\mathcal{K}$-means clustering, the performance is sensitive to the pre-set numbers of clusters as indicated from both Fig.~\ref{figa3_UEC} and Fig.~\ref{figb3_kUEC}. Given a clustering approach, the proposed optimal subchannel allocation always achieves the largest transmission rate. Compared with Fig. \ref{Fig_num_UE200},  Fig.~\ref{Fig_num_UEcluster200} further confirms that the proposed clustering with optimal subchannel allocation can well strike a balance between the distributed massive MIMO link costs and cell-free network performance.

\begin{figure}
     \centering
    \subfigure[Cutoff frequency $f_{\rm co}=100$GHz.]{
         \centering
         \includegraphics[width=3.0 in,height=2.3 in]{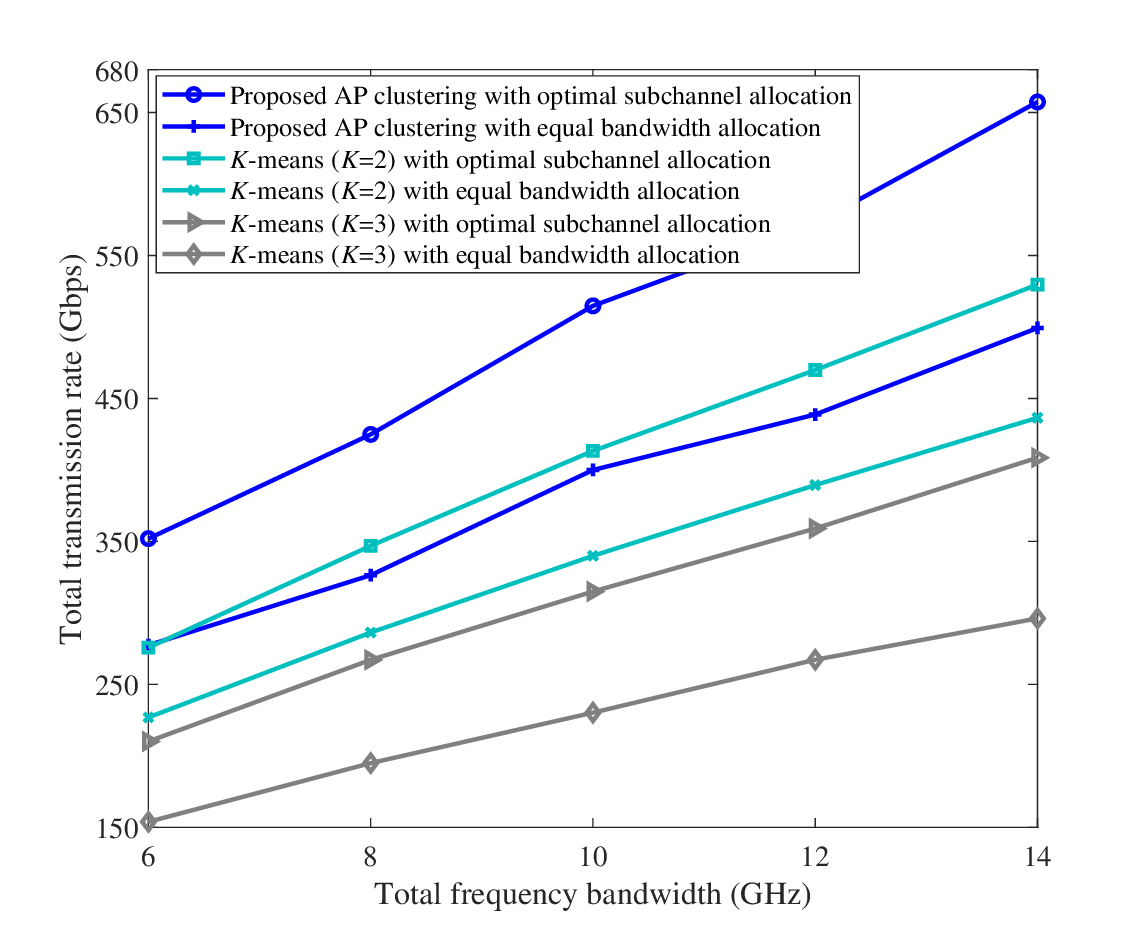}
      \label{figa3_BW}
     }
     \subfigure[Cutoff frequency $f_{\rm co}=200$GHz.]{
         \centering
         \includegraphics[width=3.0 in,height=2.3 in]{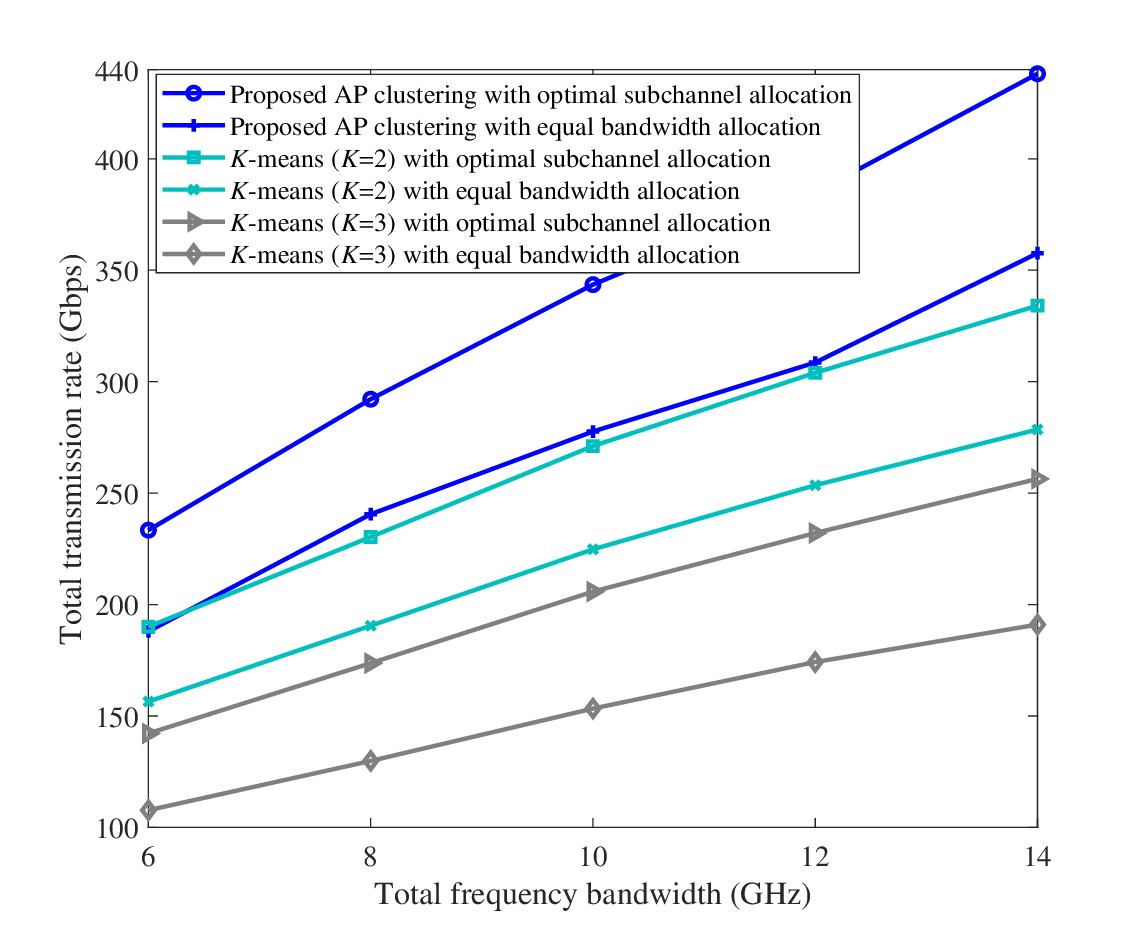}
      \label{figb3_BW}
    }
 \caption{The total transmission rate versus different levels of total frequency bandwidths  with $M=64$ and ten UEs.}
 \label{Fig_num_BWcluster200}
\end{figure}

Fig.~\ref{Fig_num_BWcluster200} shows the  total transmission rate versus different levels of total frequency bandwidths. The large frequency bandwidths at sub-THz bands enable much higher transmission rate, and the proposed  AP clustering with optimal subchannel allocation has the larger spectral efficiency than other solutions since the diversity of multiuser distributed massive MIMO can be more efficiently exploited. When the clusters are fixed, the proposed subchannel allocation can significantly improve the transmission rate since subchannels with better spectral efficiency are discovered. As seen in Fig. \ref{figb3_BW}, the advantage of the proposed hierarchical clustering solution becomes more apparent in the higher-frequencies.

\section{Conclusion}
Large-scalable cell-free network with leaky-wave antennas  is a scalable and sustainable network architecture, to exploit the advantages of both cell-free massive MIMO and spatial-spectral coupling effects. To the best of our knowledge, this is the first work to have established the large-scale cell-free THz networks with spatial-spectral coupling awareness, where the subchannel allocation, initial access and cluster formation mechanisms have been comprehensively studied. A novel cross-entropy based subchannel allocation solution was designed to maximize the total transmission rate in the considered large-scale cell-free system. Then, a joint initial access and cluster formation approach was developed to address the effects of both AP positions and AP-UE links without the need for fixing the number of clusters, to enhance the scalability of the large-scale cell-free networks. Moreover, to mitigate the inter-cluster interference, a subchannel allocation was provided to maximize the transmission rate while keeping the minimum QoS level of each cluster. The results demonstrated that besides the subchannel allocation, AP clustering could be a big factor in the large-scale cell-free networks at sub-THz bands. The proposed solutions performed much better than the state of the art methods and struck a balance  between the distributed massive MIMO link budget and cell-free network performance.

\bibliographystyle{IEEEtran}

\end{document}